\documentclass{Interspeech}

\interspeechcameraready

\title{A Dataset for Automatic Assessment of TTS Quality in Spanish}

\author[affiliation={1,2}]{Alejandro Sosa}{Welford}
\author[affiliation={3,4}]{Leonardo}{Pepino}

\affiliation{Ingeniería de Sonido}{Universidad Nacional de Tres de Febrero}{Argentina}
\affiliation{Cognitive Neuroscience Center}{Universidad de San Andrés}{Argentina}
\affiliation{Instituto de Investigacion en Ciencias de la Computación (ICC)}{CONICET-UBA}{Argentina}
\affiliation{Departamento de Computación}{Universidad de Buenos Aires}{Argentina}
\email{alejandrososawelford@gmail.com, lpepino@dc.uba.ar,}

\usepackage{comment}

\begin{document}

\maketitle

\begin{abstract}
This work addresses the development of a database for the automatic assessment of text‐to‐speech (TTS) systems in Spanish, aiming to improve the accuracy of naturalness prediction models. The dataset consists of 4,326 audio samples from 52 different TTS systems and human voices and is, up to our knowledge, the first of its kind in Spanish. To label the audios, a subjective test was designed based on the ITU‐T Rec. P.807 standard and completed by 92 participants. Furthermore, the utility of the collected dataset was validated by training automatic naturalness prediction systems. We explored two approaches: fine-tuning an existing model originally trained for English, and training small downstream networks on top of frozen self-supervised speech models. Our models achieve a mean absolute error of 0.8 on a five-point MOS scale. Further analysis demonstrates the quality and diversity of the developed dataset, and its potential to advance TTS research in Spanish.
\end{abstract}

\section{Introduction}
In recent years, voice synthesis technologies have become increasingly integrated into everyday human activities. They are widely used in virtual voice assistants such as Siri, Gemini and Alexa, as well as in accessibility applications designed to assist visually impaired individuals. Additionally, they play a significant role in entertainment media, including audiobooks and video games.

One of the most important quality factors of a text-to-speech (TTS) system is the perceived naturalness (human-likeness) of the synthesized voice. The standard approach for this evaluation relies on subjective listening tests, typically conducted using standardized datasets, most commonly in English or Mandarin Chinese. The Mean Opinion Score (MOS) \cite{itu} is the most widely used method for this assessment. This metric ranges from 1 to 5, with real human speech typically scoring between 4.5 and 4.8 \cite{TTSsurvey}. Although subjective testing is expensive in terms of time and resources, until recently, there were no automated algorithms capable of accurately modeling this metric. Traditional quality assessment systems in telecommunications, such as PESQ \cite{pesq} and POLQA \cite{polqa}, require both a degraded signal and a high-quality reference for comparison. However, this approach is not feasible when working with generative audio. It should also be noted that these models focus on distortions caused by transmission channels and were not designed to assess speech naturalness.

To overcome these limitations, non-intrusive methods for predicting speech quality have emerged. Among the most notable approaches, NISQA \cite{nisqa} has shown promising results by training a CNN-LSTM network on a dataset of English speech samples evaluated using POLQA. Another method, NORESQA-MOS \cite{noresqa}, aims to predict a generic subjective quality score for an audio sample relative to a non-matching reference. The proposed system is formulated based on the comparison between the evaluated signal and the non-matching reference, trained in an unsupervised way. Both studies report high correlation between their predicted metrics and human evaluations and claim to be language-independent.

Despite these advancements, a major limitation remains: the lack of a dedicated Spanish-language corpus for training and evaluating naturalness prediction models. This gap hinders cross-linguistic generalization and limits the development of Spanish-specific TTS evaluation methods. To address this, we present the development of a new dataset containing a diverse set of artificial Spanish voices with varying levels of synthesized quality. During data collection, a survey was conducted, gathering 4,236 speech samples rated by 92 Spanish speaking participants.

Our main contributions are: i) The development of a database for the automatic assessment of TTS systems, which, up to our knowledge, is the first of its kind in Spanish; ii) Fine-tuning NISQA on the new dataset to develop the first automated Spanish TTS evaluation model; iii) Exploring the use of self-supervised learning (SSL) speech models for automated TTS quality assessment.
Moreover, the dataset and source codes were made publicly available.\footnote{https://github.com/asosawelford/TTS-dataset-analysis}

\section{Data Collection and Evaluation}

\subsection{Data Collection}
For data collection, we considered TTS systems spanning diverse synthesis techniques and speech quality levels (Table \ref{tab:ttssystems}). Some systems allowed selecting different speakers or controlling their characteristics, like gender, accent and regional dialect. Using these controls, we generated 40 unique speakers.

\begin{table}[ht]
\caption{TTS systems utilized (IETF BCP 47 language tags).}
\label{tab:ttssystems}
\centering
\renewcommand{\arraystretch}{1.2} 
\setlength{\tabcolsep}{4pt} 
\begin{tabular}{p{2.5cm} p{2.5cm} p{2.1cm}} 
\hline
\textbf{TTS system} & \textbf{Description} & \textbf{Dialect} \\ \hline
Polly \cite{amazonpolly}      & Proprietary model        & es-ES, es-US, es-MX        \\
Azure \cite{azure}         & Proprietary model        & es-ES, es-US, es-MX \\
Speechello \cite{speechello} & Proprietary model         & es-ES, es-US, es-MX           \\
Neurasound \cite{neurasound} & Proprietary model         & es‐AR                 \\
Loquendo \cite{loquendo}     & Concatenative      & es-ES                    \\
Dewitte TTS  \cite{thomasdewitte}  & Parametric    & es-ES                    \\
FastPitch \cite{fastpitch}   & Neural    & es-AR                 \\
DC-TTS \cite{dctts}          & Neural  & es-AR                 \\
Tacotron 2 \cite{tacotron2}  & Neural  & es-AR                 \\
SLR61 \cite{guevara-rukoz-etal-2020-crowdsourcing}       & Human speech        & es-AR                 \\
\hline
\end{tabular}
\end{table}

For each speaker, we synthesized 100 phrases randomly selected from the Argentinian Spanish speech dataset \cite{guevara-rukoz-etal-2020-crowdsourcing}. To establish a reference for maximum naturalness, we also included real samples taken from this dataset. These samples consist of conversational Spanish speech by 6 different speakers from Buenos Aires, Argentina. Additionally, to further increase variability in the collected stimuli, we applied two data augmentation techniques: Vocal Tract Length Perturbation (VTLP) and phase alteration. For VTLP we used the SpeechAugment library \cite{speechaugs}, with deformation factors selected randomly between 0.9 and 1.1 for each example, following recommendations by Jaitly et al. \cite{vtlp}. For the phase alteration, we used the Griffin-Lim (GL) algorithm to estimate a phase from the magnitude spectrogram of the original samples. This algorithm can help simulate artifacts characteristic of certain vocoders present in TTS systems, thereby increasing the variability of synthetic speech quality in the dataset. VTLP was applied to 400 speech samples (4 TTS speakers) and GL algorithm was applied to 200 samples (1 TTS speaker, and 1 human speaker).

\subsection{Subjective test}
The design of the subjective test followed the specifications outlined in the ITU-T Rec. P.807 standard \cite{itu807}.  All participants self-reported normal hearing.

Following volume calibration, participants were given the following instructions:``You are about to listen to a series of short audio clips containing different types of voices. Some of these voices may sound more human-like, while others may sound more robotic. The purpose of this test is to evaluate the quality of each voice. To do this, you will rate the voices on a 5-point naturalness scale, where 5 indicates a completely natural voice and 1 represents a completely unnatural voice.''

Participants were encouraged to evaluate at least 50 stimuli but could assess more or fewer at their discretion. This approach helped mitigate fatigue-induced bias.

To ensure response validity, we monitored response times and excluded any ratings submitted below a predefined threshold relative to the audio duration. Additionally, a participant’s responses were discarded if they rated a human voice the same as a data-augmented sample, which by design represented highly degraded synthetic speech. Based on these control measures, a total of 50 ratings were excluded.

To mitigate range equalization bias, speech samples were presented in batches of 5, allowing participants to compare them before rating. Each batch contained a diverse range of speech quality levels, chosen based on the results of a pilot study. The resulting dataset contains 4,326 ratings given by 92 participants. Most of the audios received a single rating.

\section{Model design}
\subsection{Fine-tuning  NISQA}
Based on the subjective test responses, we fine-tuned NISQA v1.0\footnote{https://github.com/gabrielmittag/NISQA} for Spanish TTS naturalness assessment. The model processes mel-spectrograms, segmented into 150 ms windows with a 10 ms hop length. These are passed through a CNN 
and then fed into a LSTM network to capture temporal dependencies and predict speech naturalness. Pretraining was conducted on the NISQA Corpus \cite{nisqacorpus}, a dataset of English speech samples evaluated using POLQA.

Our Spanish dataset was divided into training (3,139), validation (393), and test (392) sets for fine-tuning. To assess generalization, we ensured that no speakers or TTS systems overlapped between the data splits. We call this model NISQA fine-tuned.

The learning rate (0.001), optimizer (ADAM), and loss function (Mean Squared Error Loss, (MSELoss)) were selected following the methodology proposed in previous studies \cite{biasloss}, and the recommendations of the model's repository. Once fine-tuning was completed, the performance of NISQA v1.0, and NISQA fine-tuned was compared by calculating Pearson Correlation Coefficient (PCC), Mean Absolute Error (MAE), and Root Mean Squared Error (RMSE) on the test dataset.

\subsection{Naturalness Prediction with DenseMOS}
We explored leveraging existing speech SSL models for the task of automatic assessment of TTS quality. These models have shown strong performance across various speech-related tasks, including automatic speech recognition (ASR), speaker verification, and speech emotion recognition \cite{wav2vecASR, sslbenchmark, macary2020useselfsupervisedpretrainedacoustic}.

The first layers of wav2vec 2.0 consist of a CNN and turn the input waveform \(x\) into a sequence \(z\) of 50 features per second. Continuous segments of \(z\) are masked, and a transformer encoder processes it. At the same time, the unmasked sequence \(z\) is quantized by the model, leading to \(q\). Finally, a contrastive loss is optimized, encouraging the model to identify the quantized representation \(q\) from distractors sampled from \(q\) and belonging to other masked time-steps. This pretext task, where the model has to recover lost information from the input waveform, allows wav2vec 2.0 to learn representations that generalize across multiple downstream speech tasks.
Throughout our experiments, we used the wav2vec 2.0 base model \footnote{https://huggingface.co/facebook/wav2vec2-base} and its version fine-tuned in LibriSpeech for ASR (wav2vec2-base-960h)\footnote{https://huggingface.co/facebook/wav2vec2-base-960h}. These models were developed and open-sourced by Meta AI.

A downstream model, which we will refer to as DenseMOS, is proposed as a predictor of synthesized speech naturalness. This neural network is based on previous works \cite{pepino_emotion, boigne} that use wav2vec 2.0 speech representations for different speech processing tasks. The relatively simple architecture of this network reduces the likelihood of overfitting to the collected dataset.

As input, DenseMOS receives the wav2vec 2.0 embeddings extracted from each audio sample, which consists of the CNN encoder and the 12 transformer blocks outputs. These activations are averaged in the time axis leading to 13 vectors of 768 features each. Then, these 13 vectors are combined into a single one by performing a weighted average, where the weights \(\alpha_i\) are learned during training. Referring to each of the wav2vec 2.0 layer activations as \(f_i\), with \(f_0\) corresponding to the local encoder output, \(f_{1}\) through \(f_{12}\) to the outputs of each internal transformer block
, the activations are combined as follows:

\begin{equation}
     f = \frac{\sum_{i=1}^{N} |\alpha_i|f_i}{\sum_{i=1}^{N}|\alpha_i|}
\end{equation}

The aggregated representation is then passed through two fully connected layers with ReLU activation and dropout. A final linear layer outputs a value between 0 and 1, which is later scaled to match the MOS range.

Model hyperparameters were empirically tuned, starting from values recommended in the literature \cite{Goodfellow-et-al-2016}. The fully connected layer size was set to 128 neurons, and a dropout probability of 0.6 was used, as lower values led to overfitting. We used MSELoss, with ADAM optimizer. Two different learning rates were used: 0.001 for \(\alpha_i\) layer weights and 0.0001 for the remaining MLP layers. L2 regularization was tested but discarded, as it did not yield significant improvements. Model training was monitored using validation loss, with early stopping applied if no improvement was observed for 40 epochs.

DenseMOS was trained using wav2vec2-base embeddings, while an alternative version, referred to as DenseMOS-960h, was trained on embeddings extracted from wav2vec2-base-960h. After training, predictions were made on the test dataset, and PCC, MAE, and RMSE were computed for both models.

All model training was conducted on a GeForce RTX 3060 (6 GB). Fine-tuning NISQA took approximately 30 minutes, while DenseMOS models required 10 minutes for training. The total parameter count for NISQA fine-tuned was 304,539, while DenseMOS models contained 115,086 parameters (not counting wav2vec 2.0 embedding extraction).

\section{Results and discussion}

\subsection{Dataset meta-analysis}
A total of 4,326 audio samples were rated in the subjective survey, amounting to 226 minutes of speech data. Each speaker is associated with a gender (male or female) and a dialect. The most represented dialect is Rioplatense Spanish with 23 voices, followed by Castilian Spanish (9 voices) and Central American Spanish (7 voices). The dataset contains an equal distribution of male and female speakers, with 26 voices per gender. On average, each sample is 3.49 seconds long (SD = 1.54) and contains 9.27 words (SD = 4.02). Label distribution can be found alongside the source code and dataset.

A total of 92 participants took part in the study, 90 of whom were Argentine, while the remaining 2 were Spanish. The average age was 29.9 years (SD = 8.2). There were 64 male participants, 25 female participants, and 3 who preferred not to specify their gender. Participants also reported their degree of familiarity with virtual assistants and artificial voices, rated on a scale from 1 to 5. The average familiarity level was 3.00 (SD = 1.30).

\begin{figure*}[ht]
    \centering
    \includegraphics[width=1\textwidth]{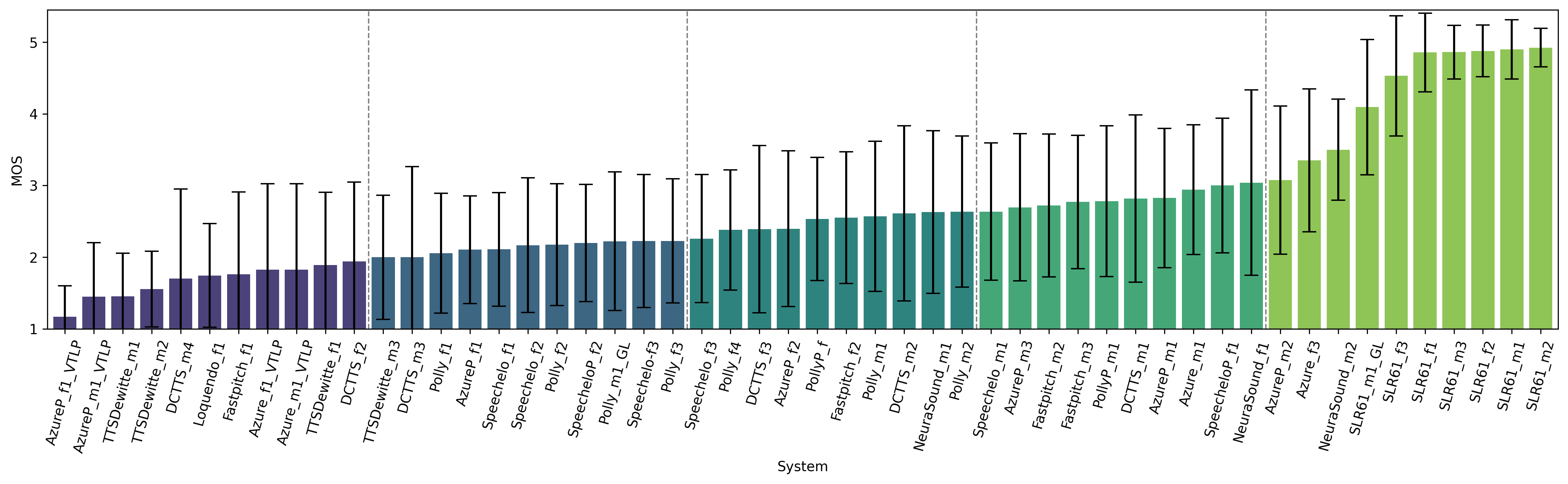}
    \caption{MOS received for each evaluated system in the subjective test. The model names correspond to the ones presented in Table \ref{tab:ttssystems}. If multiple tiers of quality were available for a given model, the letter \textbf{P} indicates the version with the advertised higher quality. Gender and voice number are indicated, and VTLP and GL suffixes specify if the samples were generated through data augmentation.}
    \label{fig:puntajes_promedio}
\end{figure*}

By taking the average score received by the speech samples of each evaluated system, we can plot their respective MOS (Figure \ref{fig:puntajes_promedio}). As expected, real human voices received the highest MOS scores, followed by various neural network-based TTS systems. Notably, different voices within the same company and technology achieved varying score ranges. This suggests that certain voices may be preferred over others, regardless of the overall quality of the TTS model.
Additionally, in many cases, the standard deviation of the received scores is relatively high. This variability could result from a combination of factors like participants having different rating strategies, with some being more critical and others more lenient in their assessments, and TTS models producing speech of varying quality depending on the input text.

To measure inter-rater agreement, we computed two metrics: Intraclass Correlation Coefficient (ICC) using a two-way random-effects model, as well as Krippendorff’s Alpha. Since most samples received only a single rating, we grouped them based on their originating TTS system or real human speaker. For ICC, missing values were imputed using the mean score of the corresponding system, while Krippendorff’s Alpha inherently supports missing data. The resulting Krippendorff’s Alpha = 0.56 and ICC(2,1) = 0.68, indicated moderate to good inter-rater agreement, respectively.

To determine whether significant differences exist between the MOS scores of different speakers, we applied the Kruskal-Wallis H-test. This non-parametric test is used to assess differences between multiple independent groups without assuming normality in the data distribution. The test yielded an H-statistic of 2101 with a p-value \textless  0.01, indicating that there are significant differences in the scores among the evaluated systems.

To determine which specific system pairs exhibit significant differences, a post-hoc analysis was performed using Tukey’s range test.  Given that more than 50 systems were evaluated, direct pairwise comparisons would be difficult to interpret. Therefore, the systems were grouped into five bins based on their average score. The post-hoc test revealed significant differences between all proposed group pairs, with p-values below 0.05 for all comparisons. These group divisions are represented in Figure \ref{fig:puntajes_promedio} by vertical dotted lines.

\subsection{Dataset validation}
The performance of the discussed models (NISQA v1.0, NISQA fine-tuned, DenseMOS and DenseMOS-960) on the test set is presented in Table \ref{tab:nisqa_results}

\begin{table}[ht]
\centering
\begin{tabular}{cccc}
\hline
\textbf{}         & \textbf{PCC $\uparrow$} & \textbf{MAE $\downarrow$} & \textbf{RMSE $\downarrow$} \\ \hline
NISQA  (v1.0)      & 0,71         & 0,99       & 1,19          \\ 
NISQA fine-tuned & \textbf{0,73}        & 0,81         & \textbf{1,06}          \\ 
DenseMOS      & 0,62         & 0,81        & 1,07         \\ 
DenseMOS-960h      & 0,60         & \textbf{0,80}       & 1,07         \\ \hline
\end{tabular}

\vspace{1em}
\caption{NISQA and DenseMOS model performance on test set. (95\% confidence intervals) } 
\label{tab:nisqa_results}
\end{table}
\vspace{-1em}
Fine-tuning NISQA on the collected dataset resulted in a performance improvement over the original model (v1.0). PCC improved from 0.71 to 0.73, while the MAE decreased from 0.99 to 0.81. These results are in line with those reported by Mittag in Section 4 of his study \cite{nisqa}, where fine-tuned versions of NISQA v1.0 achieved correlation improvements from 0.71 to 0.77 on datasets from the BLIZZARD and VCC competitions. The similarity between these findings suggests that naturalness prediction on our dataset aligns with other well-established benchmarks.
Both DenseMOS models achieved performance comparable to NISQA Fine-Tuned. The models based on wav2vec 2.0 representations yielded lower correlation values, while their MAE remained approximately the same.

\begin{figure}[ht]
    \centering
    \includegraphics[width=1\linewidth]{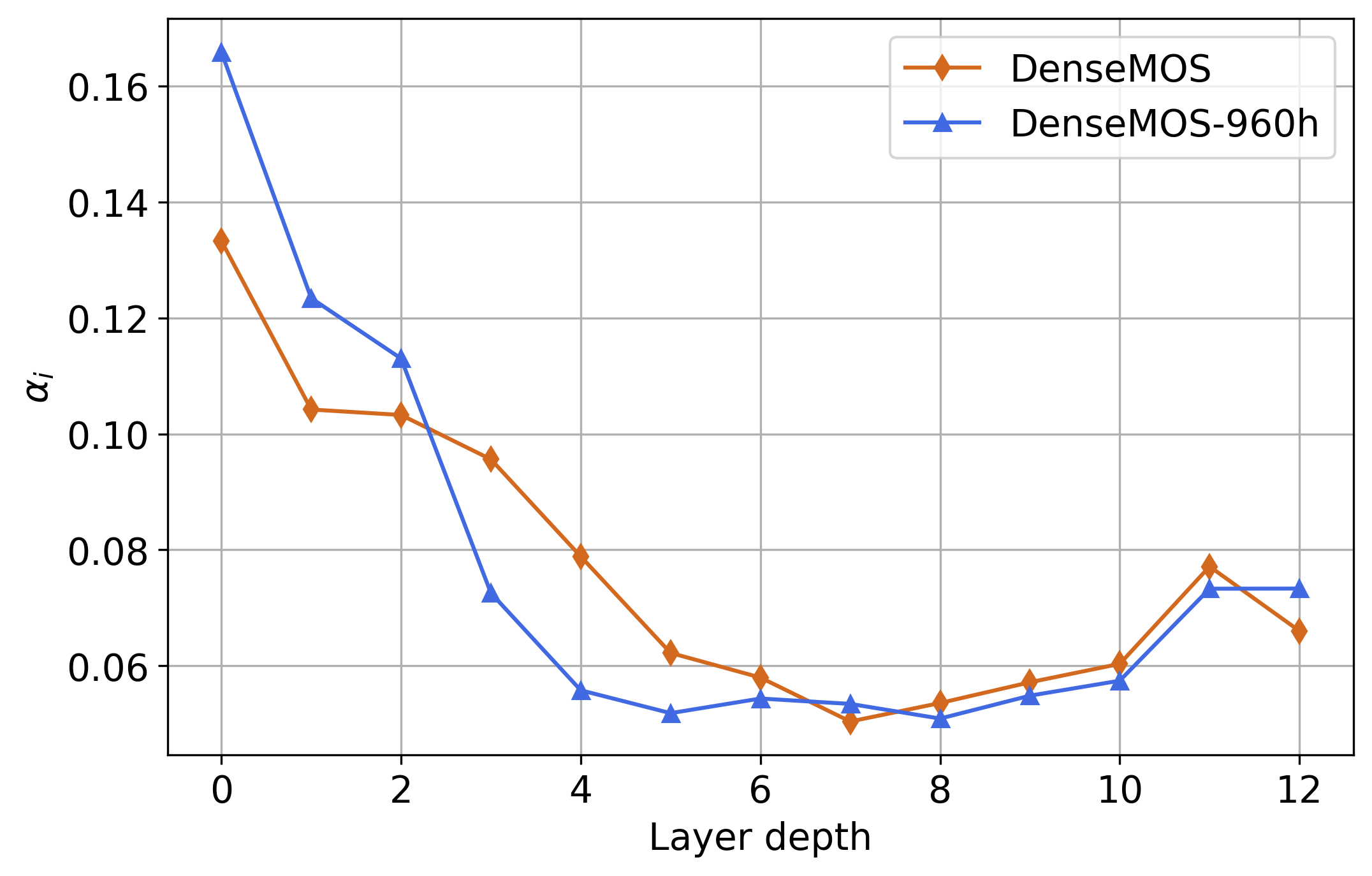}
    \caption{Weights of the trainable weighted average layer of DenseMOS and DenseMOS-960h models}
    \label{fig:layers}
\end{figure}    

The learned weights of the averaging layer in each neural network are presented in Figure \ref{fig:layers}. Index 0 corresponds to the output of the CNN encoder in wav2vec 2.0, while indices 1 to 12 represent the outputs of each transformer block.
In both cases, the highest weight was assigned to the CNN encoder output, which contains information within 25 ms intervals. The following layer activations have decreasing weights, which suggests that the task benefits primarily from local low-level representations, rather than from the more abstract contextualized features encoded in the deeper transformer layers.
The lower weights assigned to the intermediate and final layers may be explained by the fact that as we move deeper into the transformer blocks, the learned representations become increasingly aligned with the original training and fine-tuning tasks of wav2vec 2.0, which were primarily designed for ASR. 
Despite their architectural differences, NISQA and DenseMOS achieved comparable performance, suggesting that CNN-based architectures are well-suited for capturing the key features necessary for naturalness prediction. This is further backed by the highest layer weights assigned to the wav2vec 2.0 CNN encoder output.

The similarity in the results also suggests that prediction performance may be constrained by the nature of the dataset itself.  One contributing factor is the uneven distribution of MOS labels, particularly in the upper rating range. Notably, MOS scores around 4 are underrepresented, as seen in Figure \ref{fig:puntajes_promedio}.

\section{Conclusions}
This study focused on the development and validation of a dataset for the automatic evaluation of TTS systems in Spanish. The collected dataset consists of 4,326 audio samples from 52 different speakers (12 different TTS systems and 6 human voices). The samples span a wide range of speech quality, dialects, and speaker genders. A subjective evaluation was conducted following the ITU-T Rec. P.807 recommendation, with 92 participants providing MOS ratings. These responses served as a valuable reference for training and validating automatic naturalness prediction models.

In order to evaluate the usefulness of the developed dataset, NISQA was fine-tuned using the collected data, improving its speech naturalness prediction performance (achieving a PCC of 0.73 and an MAE of 0.81) on the test dataset and outperforming the baseline NISQA v1.0 model. Additionally, downstream models were trained from scratch using wav2vec 2.0 representations. These networks demonstrated comparable performance to NISQA fine-tuned, with a PCC of 0.62 and a MAE of 0.80.

Certain limitations were identified in model predictions. Specifically, all models struggled to accurately predict MOS values near 4, likely due to the underrepresentation of such scores in the dataset. Additionally, the models exhibited a bias toward predicting the dataset’s mean MOS score, which may be mitigated by exploring alternative loss functions.
These findings suggest that further exploration is needed regarding the design of naturalness prediction models.

The results of this study demonstrate that the collected dataset is a valuable resource for training and evaluating naturalness prediction models in Spanish. The achieved MAE and RMSE values are comparable to those reported in previous studies, supporting the validity and applicability of this dataset for future research.


\bibliographystyle{IEEEtran}
\bibliography{template}

\end{document}